\documentclass[cmp]{svjourmod} 

\usepackage{xspace}
\usepackage{amssymb,amsfonts,amsfonts}
\newcommand{\bd}{\begin{definition}}                
\newcommand{\ed}{\end{definition}}                  
\newcommand{\bc}{\begin{corollary}}                 
\newcommand{\ec}{\end{corollary}}                   
\newcommand{\bl}{\begin{lemma}}                     
\newcommand{\el}{\end{lemma}}                       
\newcommand{\bp}{\begin{proposition}}            
\newcommand{\ep}{\end{proposition}}                
\newcommand{\bere}{\begin{remark}}                  
\newcommand{\ere}{\end{remark}}                     

\newcommand{\bt}{\begin{theorem}}
\newcommand{\et}{\end{theorem}}

\newcommand{\be}{\begin{equation}}
\newcommand{\ee}{\end{equation}}

\newcommand{\bit}{\begin{itemize}}
\newcommand{\eit}{\end{itemize}}

\newcommand{\N}{\ensuremath{\mathbb{N}}\xspace}     
\newcommand{\R}{\ensuremath{\mathbb{R}}\xspace}     

\newcommand{\ptl}{\ensuremath{\omega}\xspace}
\newcommand{\eh}{\ensuremath{F}\xspace}

\newcommand{\dd}{{\rm d}}
\newcommand{\p}{\partial}

\begin{document}

\title{Connecting solutions of the Lorentz force equation do exist}



\author{E. Minguzzi\inst{1} \fnmsep \inst{2} \and M. S\'anchez\inst{3}}
\institute{Departamento de Matem\'aticas, Plaza de la Merced 1--4,
 E-37008 Salamanca, Spain \\ \email{minguzzi@usal.es}, 
 \and INFN, Piazza dei Caprettari 70,
I-00186 Roma, Italy  \and Departamento de Geometr\'{\i}a y Topolog\'{\i}a
Facultad de
Ciencias, Avda. Fuentenueva s/n. E-18071 Granada, Spain\\
\email{sanchezm@ugr.es}  }
\authorrunning{E. Minguzzi \and M. S\'anchez}




\date{}
\maketitle

\begin{abstract}
Recent results on the maximization of the charged-particle action
$I_{x_0,x_1}$ in a globally hyperbolic spacetime are discussed and
generalized. We focus on the maximization of $I_{x_0,x_1}$ over a
given causal homotopy class ${\cal C}$ of curves connecting two
causally related events $x_0  \leq x_1$. Action $I_{x_0,x_1}$ is
proved to admit a maximum  on ${\cal C}$, and also one in the adherence of each timelike homotopy class $C$.  Moreover, the maximum $\sigma_0$
on ${\cal C}$ is {\em timelike} if ${\cal C}$  contains a timelike curve (and the degree of differentiability of all the elements is at least $C^2$).

In particular, this last result yields a complete Avez-Seifert type solution
to the problem of connectedness through trajectories of charged
particles in a globally hyperbolic spacetime endowed with an {\em
exact} electromagnetic field: fixed any charge-to-mass ratio
$q/m$, any two chronologically related events $x_0 \ll x_1$ can be
connected by means of a timelike solution of the Lorentz force
equation (LFE) corresponding to $q/m$. The accuracy of
the approach is stressed by many examples, including an explicit counterexample
(valid for all $q/m$) in the non-exact case.


As a relevant previous step, new properties of the causal path
space, causal homotopy classes and cut points on lightlike
geodesics are studied.

\end{abstract}

\newpage

{
\tableofcontents

\contentsline {section}{ $\quad \, $Appendix: causality of
Kaluza-Klein metrics }{\pageref{sa1}} }

\section{Introduction}
Recently, there has been a renewed interest in the existence of
solutions to the Lorentz force equation (LFE; equation
(\ref{lorentz}) below) connecting two events $x_0 \ll x_1$ of a
spacetime $M$, for an (exact) electromagnetic field, $F=d\omega$.
Even though many, sometimes competitive, results have been
obtained
\cite{bartolo99,caponio02b,caponio02,caponio03,caponio04,minguzzi03b,minguzzi04b},
and related mathematical problems studied
\cite{bartolo01,caponio04c,mirenghi02}, the full answer to the
original problem has remained open:

\begin{quote}
{\bf Question (Q):} Assume that $M$ is globally hyperbolic, fix a
charge-to-mass ratio $q/m$, and let $x_0 \ll x_1$ be any two fixed
chronologically related events: must a  timelike solution of the
corresponding LFE, connecting
the two events, exist?
\end{quote}
Our main aim is to give a complete (affirmative) answer to this
question. Moreover, we will study also some properties of causal
homotopy classes not only essential for (Q) but also interesting
by themselves.

\smallskip
\smallskip

\noindent The existence of a length-maximizing causal geodesic
$\sigma_0$ connecting $x_0$ and $x_1$, when $x_0\leq x_1$, is a
well-known property since the works by Avez \cite{avez63} and
Seifert \cite{seifert67} (see Subsection \ref{ss31}).
For the LFE, one must solve two problems: \bit
\item[(A)]  Prove that the associated action functional $I_{x_0,x_1}$ admits local maximizers (or at least critical points).
\item[(B)] Show that one of these maximizers is timelike (at all the points).
\eit
As a difference with  Avez-Seifert geodesic case, now this second
point is not trivial and becomes essential because, otherwise, the
maximizer cannot be interpreted as a solution to the LFE (see Section \ref{s3}). Our
progress can be summarized as follows:


\begin{itemize}
\item[(A)] The existence of a maximizer of $I_{x_0,x_1}$ on each
causal homotopy class ${\cal C}_{x_0,x_1}$ will be proved. In
principle, a possible proof would  follow steps analogous to
Avez-Seifert theorem, but new technicalities would appear (to
prove the upper semi-continuity of $I_{x_0,x_1}$, or to try to
reduce the space of connecting curves to a finite-dimensional one,
as in Subsection \ref{ss23}), which propagate to problem (B).
Instead, we follow an approach based on Kaluza-Klein metrics as in
\cite{caponio03,minguzzi03b}.  This approach reduces problem (A)
to a problem on lightlike geodesics,
and gives a simple geometrical interpretation for the functional
$I_{x_0,x_1}$, which is somewhat reminiscent of the time of
arrival functional in the Fermat principle of General Relativity
\cite{kovner90,perlick90} (see Subsection \ref{ss42}). 
In the present paper, this
approach will be refined and clarified to obtain a maximizer of
$I_{x_0,x_1}$ on each causal homotopy class ${\cal C}_{x_0,x_1}$
(and in the closure of each timelike homotopy class $\overline {C}_{x_0,x_1}$).

\item[(B)] We must emphasize that, when the maximizer in ${\cal
C}_{x_0,x_1}$ is not timelike,  it becomes a lightlike geodesic
(Theorems \ref{mingu}, \ref{yomain}). This excludes the existence of
non-timelike maximizers for generic pairs $x_0 \ll x_1$, but
lightlike maximizers can exist for particular $x_0, x_1$ (Example \ref{exribbon}).
Nevertheless, as a main goal in the present paper, we prove that,
in this case, the causal homotopy class of a lightlike maximizer
only can contain  lightlike pregeodesics (some of them,  global
maximizers in the class), that is, {\em $x_0$ and $x_1$ are only
causally but not chronologically related by curves in the class}.
Thus, as $x_0 \ll x_1$, we can choose a causal homotopy class
which contains a timelike curve, and the required timelike
maximizer is obtained. As shown by means of an explicit example (Remark \ref{s5.2}), differentiability $C^2$ will be essential (this is the natural degree of differentiability for LFE, even though the associated variational problem make sense for $C^1$ differentiability).
\end{itemize}

\noindent Our main result is then the following:

\bt \label{tyomain} Let $(M,g)$ be a $C^2$ globally hyperbolic
spacetime, and $\eh$ be an exact 
electromagnetic field on $M$ ($\eh= \dd \ptl$ for some $C^2$ differential form). Choose $x_{1} \in J^+(x_0)$,
$x_1\neq x_0$,
and fix  any 
causal homotopy class
${\cal C}_{x_0,x_1}$.

For each charge-to-mass ratio $q/m \in \R$ there exists a
future-directed causal  curve $\sigma_0$ which connects $x_0$ and
$x_1$ and maximizes the corresponding action functional
$I_{x_0,x_1}$ on $\mathcal{C}_{x_0,x_1}$.

This maximizer $\sigma_0$ is lightlike if and only if ${\cal
C}_{x_0,x_1}$ only contains lightlike curves (necessarily
geodesics, up to reparametrizations). In this case, $\sigma_0$ is
a lightlike geodesic with no conjugate point to $x_0$ strictly
before $x_1$, and $x_1$ will be conjugate if there exists a second
curve in ${\cal C}_{x_0,x_1}$ which is not  a reparametrization of
$\sigma_0$. Otherwise, $\sigma_0$ is timelike, and its
reparametrization with respect to proper time becomes a solution
of the LFE  for the charge-to-mass ratio $q/m$.

In particular given $q/m$, if $x_{1} \in I^+(x_0)$  there exists
at least one solution to the LFE which connects $x_0$ and $x_1$.

\et

\smallskip

\noindent This work is organized as follows.

\begin{itemize}
\item In Section \ref{s2},
causal homotopy classes are studied, specially in globally
hyperbolic spacetimes.
In Subsection \ref{ss21}, the general framework is introduced, and
some new general properties are given (Theorem \ref{tch},
Corollary \ref{cch}; compare with \cite[Theorem 9.15]{beem96}). In
Subsection \ref{ss22}, we introduce the notion of {\em homotopic
cut point}, and  prove that, for lightlike geodesics, this point
becomes equivalent to the first conjugate point, Theorem
\ref{cch2}, Remark \ref{rch2}. This result will be  essential to
solve problem (B) above, in Section \ref{s5}. In Subsection
\ref{ss23}, first some properties of the standard approach for the
timelike path space of a globally hyperbolic spacetime
\cite{uhlenbeck75}, \cite[Ch. 10.2]{beem96} are shown to be
extendible to the causal case. More originally, Theorem
\ref{tdelfin} proves, in particular, that any geodesic limit of
curves contained in a single timelike or causal homotopy class, also  belongs
to this same causal homotopy class; this result will turn out
essential to solve problem (A) above, in Section \ref{s4}.

\item In Section \ref{s3}, LFE and question (Q) are introduced (Subsection \ref{ss31}), and the
variational framework for the LFE  is discussed (Subsection
\ref{ss32}). Even though this framework is well-known, we discuss
it with some detail, because there are some related variational
frameworks (widely studied in recent references) which may lead to
confusion. In Subsection \ref{ss33}, we summarize the known
results, and give a counterexample which shows that the basic
question (Q) was still open, in general.

\item In Section \ref{s4}, problem (A) on the existence of
local maximizers is solved, as explained above. Even though we
follow the approach in \cite{caponio03,minguzzi03b}, the proof is
rewritten completely. In fact, apart from some simplifications,
several new technicalities appear when causal homotopy classes are
considered, in both  the Kaluza-Klein fiber bundle (Subsection
\ref{ss41}) and the limit process on curves of the homotopy class
(Subsection \ref{ss42}, Lemma \ref{l4.4}). The main result, Theorem \ref{yomain}, includes a refinement on the existence of local maximizers for  the action $I_{x_0,x_1}$ in the closure of any timelike homotopy class (which remains valid for $C^1$ elements).

\item In Section \ref{s5}, problem (B) is solved completely. In Subsection \ref{s5.1}, we give a
result on the impossibility for a lightlike geodesic with
conjugate points to be a local maximizer of actions as those for
the LFE (Lemma \ref{yolema}, Remark \ref{r5.2}). Then, the timelike
character of the maximizer follows from the properties of causal homotopy classes in Section \ref{s2}. In Subsection \ref{s5.2}, we give examples which show the accuracy of our results: (i) even though there are maximizers in the closure of any timelike homotopy class, these maximizers can be lightlike in some of these classes, and one can ensure the existence of a timelike maximizer only in the whole causal homotopy class (which may contain more than one timelike homotopy classes), but (ii) if the degree of differentiability were only $C^1$, such a timelike maximizer may not exist.

\item In Section \ref{nex} we provide an example which shows that
the results obtained do not admit further generalizations to the
non-exact electromagnetic field case. Remarkably, in this example:
(a) for a suitably chosen pair $x_0 \ll x_1$,  no connecting
solution of the LFE exists, whatever the value of $q/m$ is chosen
(in particular, no timelike connecting solution exists for the
related Eq. (\ref{lorentz2})
 considered in Subsection \ref{ss32}), and (b) even though $F$
is non-exact, it is the curvature of a suitable bundle of fiber
$S^1$.

\item In Section \ref{s6} we give the conclusions. Finally, in a short
appendix a general result on global hyperbolicity for Kaluza-Klein
metrics is given. This result makes our paper self-contained
(compare with \cite{caponio04d} and \cite[Lemma 5]{caponio03}),
and it is provided not only for completeness, but also to
incorporate the recent progress on the splitting of globally
hyperbolic spacetimes in \cite{bernal03,bernal04}. In particular,
this rules out the problem of differentiability in the
parametrizations of the curves in the causal path space
\cite[Lemma 10.34]{beem96}.
\end{itemize}

\section{Causal homotopy classes} \label{s2}

\subsection{General properties} \label{ss21}

Throughout this paper, $(M,g)$ will denote a $C^{r_0}$  spacetime
(connected, time-oriented Lorentzian manifold), $r_0\in \{2,
\dots, \infty\}$  of arbitrary dimension $n_0\geq 2$ and signature
$(+,-,\dots,-)$. Nevertheless, for the  maximization result
of the action functional in Section \ref{s4}, it is enough $r_0=1$.
Without loss of generality, causal curves will be regarded as
piecewise $C^{r_0}$ (piecewise smooth) and future-directed from
now on\footnote{The reader can check that no more generality is obtained if the causal curves (in our at least $C^2$ spacetime) are regarded only as piecewise $C^1$ smooth. Even more, essentially the results in the present section can be extended if future-directed causal curves are regarded only as $C^0$ (i.e., a continuous
curve $\lambda \rightarrow \gamma(\lambda)$ which satisfies: for
any open connected subset $U\subset M$, if $\lambda <\lambda'$ and
$[\lambda, \lambda'] \subset \gamma^{-1}(U)$  then
$\gamma(\lambda)$ and $\gamma(\lambda')$ can be joined by a
piecewise smooth future directed causal curve contained in $U$;
such causal curves satisfy a Lipschitzian condition, see \cite[p.
17]{penrose72}). For example, Lemma \ref{lch} (and, then, Theorem
\ref{tch} and Corollary \ref{cch}) or Theorem \ref{cch2} hold
obviously if the longitudinal curves of the causal homotopies are
allowed to be only continuous causal curves.}. Notice that the
curves are regarded as parametrized (we will not be specifically
interested in the space of all the unparametrized causal curves,
in the spirit of Morse theory), and suitable reparametrizations
will be chosen, if necessary. In principle, strong causality is
our (minimal) ambient  causal assumption for $(M,g)$, always assumed implicitly. But most of
the results need global hyperbolicity, and sometimes this
assumption will be imposed (explicitly in this section) for simplicity. For background results
in this section, see for example \cite[Ch. 9, 10]{beem96} and
\cite[Ch. 10]{oneill83}; timelike homotopy classes have also been
studied in different contexts (see for example,
\cite{smith67,galloway84,sanchez05} and references therein); some
of the difficulties circumvented in our approach are illustrated
in the beginning of the proof of \cite[Th. 6.5]{penrose72}.

Two 
causal (resp. timelike) curves, $\gamma_i:
[\lambda_0, \lambda_1]\rightarrow M$, $i= 0,1$ with fixed extremes
     $x_j= \gamma_0(\lambda_j)=\gamma_1(\lambda_j)$, $j = 0,1$, are
{\em causally homotopic} if there exists a
{\em causal} (resp. {\em timelike}) homotopy (with fixed extremes $x_0, x_1$) connecting
$\gamma_0, \gamma_1$, i.e., a continuous map
$$
\begin{array}{rcl}
H:[0,1]\times [\lambda_0,\lambda_1] & \rightarrow & M \\
(\epsilon , \lambda ) & \rightarrow & \gamma_\epsilon(\lambda )
\end{array}
$$
such that each longitudinal curve $\gamma_\epsilon$ is causal (resp. timelike)
for all $\epsilon$.
This divides the set of 
causal curves joining $x_0$ and
$x_1$ into 
{\em causal homotopy classes}, each one containing none, one or more classes
of timelike homotopy (see Example \ref{ejecut}).
In our notation a generic causal (resp. timelike) homotopy class
will be denoted with $\mathcal{C}_{x_0,x_1}$ (resp.
$C_{x_0,x_1}$).

The following lemma to Theorem \ref{tch} solves the technical difficulty
associated to our choice of parametrized curves.
\bl \label{lch} Consider two  causally homotopic lightlike
geodesics 
$\gamma_0,\gamma_1\! : \! [\lambda_0,\lambda_1]\rightarrow M$
joining $x_0$ with $x_1$.

If $\gamma_0$ and $\gamma_1$ maximize the time-separation (or ``length'') in its causal
homotopy class ${\cal C}$ (that is, there is no timelike curve in
${\cal C}$), then  there exist a causal homotopy connecting them through longitudinal geodesics.

\el

\begin{proof}
By the hypothesis, all the longitudinal curves of the causal
homotopy $H(\epsilon,\lambda)$ between $\gamma_0$ and $\gamma_1$
are necessarily lightlike pregeodesics \cite[Proposition
10.46]{oneill83}. Let $\gamma_\epsilon:
[\lambda_0,\lambda_1]\rightarrow M $ be the (unique)
reparametrization as a geodesic of the longitudinal curve
corresponding to $\epsilon \in [0,1]$, and put $u_\epsilon =
\gamma'_\epsilon(\lambda_{0})$. The map
$$h: (\epsilon , \lambda ) \rightarrow  \gamma_\epsilon(\lambda )$$
will be continuous (and, thus, the required 
causal homotopy) if  and only if the map
$$\epsilon \rightarrow u_\epsilon, \quad \epsilon\in [0,1]$$ is continuous.
Consider the tangent space at $\gamma(\lambda_{0})$ and  the
canonical projection on the projective space $\pi: TM_{x_{0}} \to
PTM_{x_0}$. The curve of directions $\epsilon \to
\pi(u_{\epsilon})$ is continuous since $\pi(u_{\epsilon})=\pi(
\exp_{x_{0}}^{-1}H(\epsilon,\lambda))$, in any normal neighborhood
of $x_0$, independently of $\lambda$. Since $\exp_{x_{0}}
(u_{\epsilon}(\lambda_{1}-\lambda_{0}))=x_{1}$, the continuity of
the curve $\epsilon \to u_{\epsilon}$ follows directly from
\cite[Lemma 9.25 ]{beem96} (the strong causality of $M$ is used
there). \qed \end{proof}

\smallskip


\bt \label{tch} Let $\gamma_0 :
[\lambda_0,\lambda_1]\rightarrow M $ be a
lightlike geodesic which connect two fixed points $x_0, x_1$ and
maximizes the time-separation in its causal homotopy class.
If there exists a distinct geodesic $\gamma_1$ in this class then
$x_1$ is the first conjugate point of $x_0$ along $\gamma_0$ (and, then, along $\gamma_1$). \et

\begin{proof} By previous lemma, there exists a 
causal homotopy from $\gamma_0$ to $\gamma_1$ through lightlike
geodesics. Thus, exp$_{x_0}$ cannot be injective in any
neighborhood of $(\lambda_1-\lambda_0) \gamma_i'(\lambda_0),
i=0,1$, and $x_1$ is a conjugate point. Even more, it is the first
one because, otherwise, $\gamma_0$ would not maximize in its
causal homotopy class. \qed \end{proof}

\bere \label{r2.4} { The converse implication may not hold: even
though a variation of $\gamma_0$ through lightlike geodesics with
variational vector field zero at the extremes will exist
\cite[Corollary 10.40]{oneill83}, conjugate points are only
``almost meeting points'' of geodesics, that is, the longitudinal
geodesics of the variation may not reach $x_1$ }. \ere

\smallskip

\noindent Cut points on causal geodesics have been widely studied
\cite[Ch. 9]{beem96}; recall that  a lightlike geodesic ray
maximizes the time-separation  until its cut point. \bc
\label{cch} Let $\gamma: [0,b)\rightarrow M $ be a lightlike
geodesic with a cut point $\gamma(\lambda_c)$, $\lambda_c
\in(0,b)$. If $\gamma(\lambda_c)$ is not a conjugate point, then:

(1) No other lightlike geodesic which connects $\gamma(0)$ and
$\gamma(\lambda_c)$ is causally homotopic to $\gamma$.

(2) If $(M,g)$ is globally hyperbolic, there exist at least another lightlike geodesic $\hat \gamma$
(necessarily non-causally homotopic to $\gamma$) which connects
$\gamma(0)$ and $\gamma(\lambda_c)$.

\ec \begin{proof} Assertion (1) is straightforward from Theorem
\ref{tch}. Then, (2) is a consequence of the well-known existence
of a second connecting lightlike geodesic for any non-conjugate
cut point on a lightlike geodesic (see \cite[Theorem
9.15]{beem96}). \qed \end{proof}

\bere { The well-known behaviour of lightlike geodesics of
bidimensional de Sitter spacetime illustrates Corollary \ref{cch}.
No lightlike geodesic can have a conjugate point (because of
dimension 2) but any such geodesic has a cut point, reached by a
non-causally homotopic lightlike geodesic. By removing a point of
this second geodesic, the necessity of the assumption of global
hyperbolicity for (2) is stressed. }\ere

\subsection{Homotopic cut points} \label{ss22}

\noindent  For the following crucial result, we introduce an
auxiliary concept:

\bd Let $\gamma: [0,b)\rightarrow M $ be a lightlike geodesic. The
point $\gamma(\lambda_c)$, $\lambda_c>0$ is the {\em homotopic cut
point} along $\gamma$ of $\gamma(0)$,  if $\lambda_c$ is the first
point such that, for any $\delta>0$, the restricted curve $\gamma
|_{[0,\lambda_c+\delta]}$ does not maximize the length in its
causal homotopy class (that is, $\gamma|_{[0,\lambda_c+\delta]}$
is causally homotopic to a timelike curve). \ed

\bt \label{cch2} Let $\gamma: [0,b)\rightarrow M $ be a lightlike
geodesic in a globally hyperbolic spacetime $(M,g)$. The point $\gamma(\lambda_c)$ is the homotopic cut point
along $\gamma$ of $\gamma(0)$ if and only if it is the first
conjugate point to $\gamma(0)$.

\et \begin{proof} ($\Rightarrow$). Assume that
$\gamma(\lambda_c)$ is not the first conjugate point. As conjugate
points on causal geodesics are discrete \cite[Th. 10.77]{beem96},
we can choose $\lambda_\delta = \lambda_c+\delta >\lambda_c$ such
that no conjugate point $\gamma(\lambda)$ appears for  $\lambda\in
[0,\lambda_\delta]$. By hypothesis on $\gamma(\lambda_c)$, there
exists a timelike geodesic $\rho$ from $\gamma(0)$ to
$\gamma(\lambda_\delta)$ which is causally homotopic to $
\gamma|_{[0,\lambda_\delta]}$, and maximizes the time-separation
in its causal homotopy class. We can assume that, for such a
causal homotopy $H(\epsilon,\lambda)$, the longitudinal curves
$H_\epsilon$  are not lightlike pregeodesics close to $\gamma$
(neither, in particular, equal to $\gamma_0$); otherwise,
$\gamma(\lambda_\delta)$ would be conjugate to $\gamma(0)$ along
$\gamma$. We will need a  variation $h(\epsilon,\lambda)$ of
$\gamma|_{[0,\lambda_\delta]}$ which is piecewise $C^2$  (that is,
continuous and $C^2$ on each closed rectangle corresponding to a
suitable partition of the domain $(\epsilon, \lambda)$), and
satisfies the other technical properties of the following result,
to be proved at the end.

\bl \label{lreyes} Curve $\gamma_0= \gamma|_{[0,\lambda_\delta]}$
admits a piecewise $C^2$ causal homotopy
$$h: [0,1]\times [0,\lambda_\delta] \rightarrow M, \quad \quad
(\epsilon, \lambda )\rightarrow \gamma_\epsilon(\lambda)$$ with
fixed extremes, such that the longitudinal curves
$\gamma_n:=\gamma_{\epsilon_n}$ are causal for some sequence
$\epsilon_n \searrow 0$, and the variational vector field
$V=\partial_\epsilon|_0 \gamma_\epsilon(\lambda)$ is not
identically 0.

In particular, $V$ satisfies:

$$ V(0)=V(\lambda_\delta)=0, \quad \quad V\not\equiv 0, \quad \quad g(V,\gamma')\equiv 0 .$$

\el

\noindent We remark that, a priori, the longitudinal curves are
causal only for the sequence $\{\gamma_n\}$, that is, the
variation maybe ``non-admissible'',  in the terminology of \cite[Sect.
10.3]{beem96} (recall also Remark \ref{r2.4}).

\smallskip

\noindent  Recall that, according to \cite[Dfns. 10.47, 10.49,
10.54, 10.57, 10.59]{beem96}, $V$ induces a class $[V]$ which
belongs to the domain $\mathfrak{X}_0(\gamma_0)$ of the quotient
index form $\bar I$, i.e., the index form  defined on the
piecewise smooth sections (vanishing at the extremes) of the
quotient bundle $G(\gamma)$ defined by taking vector fields on
$\gamma$ modulo $\gamma'$. As each $\gamma_n$ is
causal\footnote{With our sign convention $(+,-, \dots, -)$,
different to \cite{beem96,oneill83}.}, then $g(\gamma'_\epsilon, \gamma'_\epsilon)
$ is non-decreasing at 0, and
$$0\leq \frac{1}{2}\left. \frac{\partial^2}{\partial \epsilon^2} \right|_0 \int_0^{\lambda_\delta}
g(\gamma'_\epsilon (\lambda), \gamma'_\epsilon (\lambda) )
d\lambda = I(V,V) =
\bar I([V],[V]),$$
(see also \cite[pp. 289-290]{oneill83}), in
contradiction with \cite[Th. 10.69]{beem96}.

$(\Leftarrow)$. Obviously, the homotopic cut point of $\gamma(0)$
must appear not beyond the first conjugate point. But from the
proved implication, it can neither appear before this point. \qed
\end{proof}

\smallskip
\begin{theopargself}
\begin{proof}[of Lemma \ref{lreyes}]  Fix a finite covering of
convex neighborhoods of $M$ which cover the image of $\gamma_0$,
and choose $0=\lambda_0 < \lambda_1 < \dots < \lambda_k <
\lambda_{k+1}= \lambda_\delta$ such that
$\gamma([\lambda_i,\lambda_{i+1}])$ is included in one of such
neighborhoods, $U_i$, for all $i=0, \dots , k$. Notice that, taken
normal coordinates on each $U_i$, the tangent space to
$\gamma_0(\lambda_i)$ is identifiable to $\R^n$. Even more, as
$$\lim_{\epsilon\rightarrow 0} H_\epsilon(\lambda_i)= \gamma_0(\lambda_i)$$
and the set of directions in $\R^n$ is compact, there exists a
sequence $\{\epsilon^i_n \} \searrow 0$ and a $C^1$-curve
$\alpha_i$ such that
$\alpha_i(\epsilon^i_n)=H_{\epsilon^i_n}(\lambda_i)$ for all $n$.
By taking each sequence $\{\epsilon^i_n\}$ as a subsequence of
$\{\epsilon^{i-1}_n\}$, we can assume that all the sequences are
equal $\epsilon_n \equiv \epsilon^i_n$. Iterating the process, $\alpha_i$ can be chosen $C^r$, for any finite $r>1$. Let us show that
$\alpha_{i_0}'(0) \neq 0$ can also be assumed for some $i_0$. Indeed, otherwise,
as the curves $H_\epsilon$ are
different to $\gamma_0$ for small $\epsilon$, we can find another
$C^2$ reparametrization  of some $\alpha_{i_0}$
such that its velocity at 0 does not vanish, and consider this new
parameter as the original transversal parameter of the homotopy
$H$.

Now, choose as variation $\gamma_\epsilon(\lambda)$ of $\gamma_0$
(for small $\epsilon$) the homotopy $h$ defined as: $\gamma_\epsilon$
is the unique broken geodesic which joins $\alpha_i(\epsilon)$ and
$\alpha_{i+1}(\epsilon)$ in $U_i$ when $\lambda$ varies between
$\lambda_i$ and $\lambda_{i+1}$. Let $v_{i}(\epsilon)$ be the
tangent vector at $\lambda_i$ of such a $\gamma_\epsilon$ restricted to
$[\lambda_i,\lambda_{i+1}]$. As each pair of  curves $\alpha_i$, $\alpha_{i+1}$ are $C^2$, the curve in $TM$ which maps each $\epsilon$ to
$v_i(\epsilon)$ is $C^2$ too (use  \cite[Lemma 5.9]{oneill83}). Thus, the homotopy $h$
can be written as 
\[
h(\epsilon,\lambda)\vert_{\lambda\in [\lambda_i,\lambda_{i+1}]}=\gamma_{\epsilon}(\lambda)\vert_{[\lambda_i,\lambda_{i+1}]}=\exp_{\alpha_{i}(\epsilon)}((\lambda-\lambda_{i})v_{i}(\epsilon))
\]
for small $\epsilon$, and all the required properties follow.
\qed \end{proof}
\end{theopargself}

\bere \label{rch2} { The definition of homotopic cut point is
obviously extendible to timelike geodesics and to geodesics in
Riemannian manifolds, but the analogous of Theorem  \ref{cch2}
would not hold, in general. In fact, it is easy to construct a
Riemannian manifold $(S, \dd l^2)$ with a closed geodesic
$c:[0,2]\rightarrow S$, $c'(0)=c'(2)$, without conjugate points,
such that its middle point $c(1)$  is the cut point,  and the two
pieces of the geodesic $c_0= c|_{[0,1]}$, and
$c_1(\lambda)=c(2-\lambda), \forall \lambda\in [0,1]$ are
homotopic with fixed extremes $p_0=c(0), p_1=c(1)$.

Concretely, let $S$ be the following surface embedded in Euclidean
space $\mathbb{R}^3$ (with natural coordinates $(x,y,z)$) and
induced metric $\dd l^{2}$. $S$ is obtained by gluing the
semi-cylinder $x^{2}+y^{2}=r^2$, $z \le 0$, with a spherical cap
of radius r, $x^2+y^2+z^2=r^2$, $z \ge 0$. The metric $\dd l^2$ is
$C^1$ but can even be made smooth by suitably redefining the cap
near the equator  (this manifold will be used in the next
examples; it can be also replaced by a paraboloid, with
straightforward modifications). The required geodesic then would
be $c(\lambda)= (r\cos \pi \lambda, r\sin \pi \lambda, -1)$, with
$p_0=(r,0,-1), p_1=(-r,0,-1)$. } \ere \noindent This
counterexample for Riemannian geodesics is extended to timelike
geodesics in the following example, which also shows the possible
existence of different timelike homotopy classes in a single
causal one.
\begin{example} \label{ejecut} {
Consider the Riemannian surface $S \ni p_0, p_1$ as above, and
define the spacetime $M= \R \times S, g=dt^2-dl^2$. Choose $T>\pi
r$, and take $z_0=(0,p_0), z_1=(T,p_1) \in M$. The timelike
geodesics $\sigma_i(\lambda)=(T \lambda, c_i(\pi \lambda)), i=0,1$
maximize the time-separation between $z_0$ and $z_1$, and do not
have conjugate points. If $T>2+ \pi r$ then they are timelike
homotopic, and if $T=2+\pi r$ then they are causally homotopic,
but not timelike homotopic (if $S$ is smoothed as suggested above,
the critical value $T=2+\pi r$ must be replaced by $T=L$ where $L$
satisfies: (i) any smooth homotopy in $S$ between $c_0$ and $c_1$
contains a curve of length $L$, (ii) no $L'>L$ satisfies (i)).
}\end{example}

\subsection{Arc-connectedness of the closure of the classes} \label{ss23}

Now, let us remark some properties of causal homotopy classes in
globally hyperbolic spacetimes, in relation to Uhlenbeck's study
of the timelike case \cite{uhlenbeck75}, carefully developed
further by Beem et al. \cite{beem96}. Our main aim is to show an
appropiate sense of compactness of the piecewise geodesics in
timelike homotopy classes (Theorem \ref{tdelfin}, Remark
\ref{rdelfin}), which will be directly extendible to the fibered
classes in Section \ref{s4}. Along this subsection, $(M,g)$ will
be globally hyperbolic, and a {\em temporal Cauchy function}
$t:M\rightarrow \R$  (see the Appendix)
is chosen to parametrize all the causal curves. Two points $x_0< x_1$ will be
also fixed and, as it is not restrictive to assume $t(z_i)=i, i=0,1$, all the
connecting causal curves will be parametrized in $[0,1]$.

Following \cite[Sect. 10.2]{beem96} (our basic reference throughout this subsection), there
exist a $N > 0$ and a partition  $0=t_0< t_1 < \dots <t_N <t_{N}=1$ of $[0,1]$ such that
 for any {\em causal chain} $(z_0,
\ldots z_N)$, $z_i \in M$, $z_{i} < z_{i+1}$ with $t(z_i)=t_i$ and
$z_0=x_0$, $z_N=x_1$, there exists one and only one maximal causal
geodesic connecting $z_i$ with $z_{i+1}$ (thus, $(z_0, \dots ,
z_N)$ can be identified with the piecewise causal geodesic
obtained connecting $z_i$ with $z_{i+1}$).
Let ${\cal M}_{x_0, x_1}$ 
be the space of such causal chains and $M_{x_0, x_1}$ the subset
containing all the chronologically related chains (i.e., $z_{i} <<
z_{i+1}$). Clearly,  ${\cal M}_{x_0, x_1}$ and $M_{x_0, x_1}$ are
subsets of a product of $N-1$ Cauchy hypersurfaces $S_i$ (the
fixed extremes of the chain can be disregarded) and, then,
inherit a  topology. 
By using that the relation $\leq$ is closed on convex
neighborhoods, it is straightforward to check: \bp $ $
\begin{itemize}
\item[(1)] The space of causal chains ${\cal M}_{x_0, x_1}$ is
compact. \item[(2)] The space of chronological chains $M_{x_0,
x_1}$ is open in $S_1 \times \dots \times S_{N-1}$, and its
closure is included in ${\cal M}_{x_0, x_1}$.
\end{itemize}
 \ep \noindent One can check, reasoning for
${\cal M}_{x_0, x_1}$ as in \cite[Proposition 10.36]{beem96} for
$M_{x_0, x_1}$,  that there exists a well defined length
non-decreasing retraction from the set ${\cal N}_{x_0,x_1}$ of all
the $t-$parametrized causal curves connecting $x_0$ and $x_1$ to
${\cal M}_{x_0,x_1}$ (the topology of ${\cal N}_{x_0,x_1}$ can be
chosen as the topology associated to the uniform distance $d$,
i.e.: $d(\gamma_1, \gamma_2)=$ Max$\{d_R(\gamma_1(t),\gamma_2(t)):
t\in [0,1]\}$, where $d_R$ is the distance associated to any
auxiliary Riemannian metric).

Given an arc-connected component $U_{x_0,x_1}$ (resp. ${\cal
U}_{x_0,x_1}$) of $M_{x_0, x_1}$ (resp. ${\cal M}_{x_0, x_1}$),
its piecewise smooth geodesics are timelike (resp. causally)
homotopic. In fact, any continuous curve $( z_0(\epsilon), \ldots
z_N(\epsilon))$ in $U_{x_0,x_1}$ (resp. ${\cal U}_{x_0,x_1}$)
joining two given chains, also yields the required homotopy
between the associated piecewise geodesics. 
We will be interested just in the problem of the invariance of the
arc-connected components under limits. More precisely, recall
that, as  $M_{x_0, x_1}$ is an open subset of a manifold, its
connected and arc-connected components are equal. Even more, the
following property (which is {\em false} for open subsets of
$\R^N$, in general) holds:

\bl \label{ldelfin} Any chain $(z_{0,\infty}, \dots ,
z_{N,\infty})$ in the closure of a (arc-)connected component
$U_{x_0,x_1}$ (resp. ${\cal U}_{x_0,x_1}$) of $M_{x_0, x_1}$
(resp. ${\cal M}_{x_0,x_1}$), can be connected to any element of
$U_{x_0,x_1}$ (resp. ${\cal U}_{x_0,x_1}$) by means of an arc
totally included in $U_{x_0,x_1}$ (resp. ${\cal U}_{x_0,x_1}$),
except one extreme.

\el
\begin{proof}
We will reason for $U_{x_0,x_1}$, being  analogous for ${\cal
U}_{x_0,x_1}$. Consider any converging sequence of chronological
chains  in $U_{x_0,x_1}$, $\{(z_{0,k}, \dots , z_{N,k})\}_k$
$\rightarrow$ $(z_{0,\infty}, \dots $ $, z_{N,\infty})$. Let
${\cal N}_i$ be the convex neighborhood which contains
$z_{i,\infty}$, $z_{i+1,\infty}$ and make lighter the
notation putting $x_k= z_{i,k}$, $y_k= z_{i+1,k}$, for $k=1,2..., \infty$. 
The result follows immediately by applying the following claim
recurrently for $i=0,\dots, N-1$:

{\em Claim:} Let $x_k<<y_k$, $k\in \N$, and $x_\infty < y_\infty$ be as above. If there
exists a continuous curve $\alpha:[0,1]\rightarrow {\cal N}_i$ and a decreasing sequence
$\lambda_k\rightarrow 0$ such that $x_k= \alpha(\lambda_k)$ for large $k$, then there exists a
continuous curve $\beta:[0,1]\rightarrow {\cal N}_i$ such that, for some $\epsilon\in(0,1]$:

$$\alpha(\lambda) << \beta(\lambda) , \quad \beta(\lambda_k)=y_k , \quad \forall \lambda, \lambda_k\in (0,\epsilon).$$

\smallskip

\noindent Notice that this claim is obvious in the particular case $x_\infty <<y_\infty$. For
the general case, let $v_k$ be the velocity at zero of the unique timelike geodesic
(causal, when $k=\infty$), defined on $[0,1]$,  from $x_k$ to $y_k$. As the bundle of the future
timelike cones on $\alpha$ is arc-connected, we can find a continuous timelike vector field $V$
on $\alpha$ such that $V(\lambda_k)= v_k$. The required curve is then
$$\beta(\lambda) = \exp_{\alpha(\lambda)}(V(\lambda)).$$
\qed \end{proof}

Recall that, in the case of  ${\cal U}_{x_0,x_1}$, the a priori
excluded extreme will also belong to ${\cal U}_{x_0,x_1}$. Summing
up: \bt \label{tdelfin} Let $(M,g)$ be globally hyperbolic.
\begin{itemize}
\item[(1)] The closure $\overline{U}_{x_0,x_1}$ of any connected
component $U_{x_0,x_1}$ of
 $M_{x_0, x_1}$ is arc-connected.
\item[(2)] Any arc-connected component ${\cal U}_{x_0,x_1}$ of ${\cal
M}_{x_0, x_1}$ is closed.
\end{itemize}
\et

\bere \label{rdelfin} {\rm In particular, when $\{\gamma_k\}_{k\in
\N }$ is a sequence of causal geodesics in the same causal (resp.
timelike) homotopy class, such that its initial velocities
converge to the velocity of a geodesic $\gamma_0$, then $\gamma_0$
belongs to the same causal homotopy class (resp. to the adherence
of the timelike homotopy class -- which is included in the same
causal homotopy class of the sequence).


} \ere

\section{Connectedness through solutions to the LFE} \label{s3}
In the following sections we shall consider three kinds of
manifolds: the spacetime $M$ of dimension $n_0 \ge 2$, the
Kaluza-Klein spacetime $P$ of dimension $n_0+1$ and  (in some
examples) a spacelike hypersurface
 $S$ of dimension $n_0-1$. We adopt the convention of denoting curves belonging to $P$ with
$\gamma$, curves belonging to $M$ with $\sigma$ or $x$, and curves
belonging to $S$ with $c$.

\subsection{Avez-Seifert type problem for the LFE}\label{ss31}
Consider  on $M$ a fixed (exact) electromagnetic field
$F=d\omega$, where $\omega$ is any differential 1-form. A point
particle of rest mass $m>0$ and electric charge $q\in \R$, moving
under $F$, has a timelike worldline which satisfies the {\em
Lorentz force equation (LFE)} (cf. \cite[section 3.1]{misner73},
\cite[section 11.9]{jackson75} or \cite[section 23]{landau62})
\begin{equation} \label{lorentz}
 D_s \left(\frac{\dd x}{\dd
s}\right)=\frac{q}{m}\hat F(x)\left[\frac{\dd x}{\dd s}\right].
\end{equation}
Here the units are such that $c=1$, $x=x(s)$ is the world line of
the particle parameterized with proper time, $\frac{\dd x}{\dd s}$
is the velocity, $D_s\left(\frac{\dd x}{\dd s}\right)$ is the
covariant derivative of $\frac{\dd x}{\dd s}$ along $x(s)$
associated to the Levi-Civita connection of $g$, and $\hat
F(x)[\cdot]$ is the linear map  on $T_x M$ defined by
\[
 g(x)[v,\hat F(x)[w]] = F(x)[v,w] ,
\]
 for any $v,\ w\in T_xM$.

We remark that $s$ must be the proper time parametrization,
otherwise the constant $q/m$ in front of $\hat{F}$ cannot be
interpreted as the charge-to-mass ratio of the particle. Recall
that, for the LFE, the ratio $q/m$ is fixed, but the individual
values of $q$ and $m$ become irrelevant\footnote{Sometimes  the mass of
the particle is assumed to be known, and the curve $x$ is
parametrized with $r=s/m$. In this case the LFE becomes equivalent
to the system \cite[definitions 3.1.1 and 3.8.1]{sachs77}
$D_r \left(\frac{\dd x}{\dd r}\right)= q\hat F(x)\left[\frac{\dd
x}{\dd r}\right]$, 
$|\frac{\dd x}{\dd r}|= m$.}.

As commented in the introduction, question (Q) becomes natural now.
For the case $q/m=0$, the LFE is just the geodesic equation, and
the solution to question (Q) is well-known \cite[Theorem 3.18,
Proposition 10.39]{beem96}, \cite[Proposition 14.19]{oneill83},
\cite[Proposition 6.7.1]{hawking73}:

\begin{theorem}\label{tas} {\rm (Avez \cite{avez63}, Seifert \cite{seifert67})}. Let $(M,g)$ be a globally
hyperbolic spacetime, and $x_0\leq x_1$ two causally related events. Then, in each causal homotopy class,
 ${\cal C}_{x_0,x_1}$
there exists a a causal geodesic $\sigma$ which maximizes the
length among the causal curves  in the class.
\end{theorem}
In particular, if $x_0 \ll x_1$ the two points can be connected by means of a timelike
geodesic. If $x_1\in E^+(x_0)=J^+(x_0)\backslash I^+(x_0)$ then $x_0$ and $x_1$ can still be
joined by a lightlike geodesic, 
but this case does not make sense for the LFE. One can
also wonder for the connectedness of $x_0, x_1$ by means of a
geodesic even if they are not causally related, as in variational frameworks described below.
Although this question has a geometrical interest
(see for instance the survey \cite{sanchez01}), it does not have a direct physical interpretation,
nor equivalence for LFE.

\subsection{Related variational problems}\label{ss32}
Question (Q) can be approached as a variational one \cite[section
16]{landau62}. Indeed, let $\mathcal{N}_{x_0,x_1}$ be the set of all
(piecewise) $C^{1}$ causal curves $\sigma: [0,1] \to M$ from $x_0$ to $x_1 > x_0$.
Using $F=d\omega$, consider the functional $I_{x_0,x_1}$ on
$\mathcal{N}_{x_0,x_1}$,
\begin{equation} \label{action}
I_{x_{0},x_{1}}[\sigma]=\int _{\sigma} (\dd s+\frac{q}{m} \omega) =
\int_0^1
\left(\sqrt{g(\sigma'(\lambda ), \sigma'(\lambda))} +
\frac{q}{m}\omega (\sigma'(\lambda))\right)d\lambda ,
\end{equation}
for all $\sigma \in \mathcal{N}_{x_0,x_1}$.
The functional is invariant under monotonic reparametrizations of
$\gamma$; in this sense, when talking about critical points, we
may refer to non-parametrized curves. The connecting timelike
solutions of the LFE (\ref{lorentz}), if they exist, are critical
points of this functional, as it follows from a computation of the
Euler-Lagrange equation. Conversely, every timelike extremal of
this functional, once parametrized with respect to proper time, is
a solution of the LFE.

In the geodesic case $q/m=0$, it is convenient to replace
functional $I_{x_0,x_1}$ by the ``energy'' functional \be
\label{energy} 
E[\sigma]= \frac{1}{2}\int_0^1 g(\sigma'(\lambda ),
\sigma'(\lambda))d\lambda \ee because of several reasons: (i) the
critical curves of this functional are parametrized directly as
geodesics, and (ii) the domain of the functional can be enlarged
to include non-causal curves (making sense also for non-causally
related $x_0, x_1$) and, then, non-causal connecting geodesics
also become critical points. Moreover,  the choice of extremes
$\lambda_0=0, \lambda_1=1$ simplifies the domain of curves without
loss of generality. Nevertheless, if one only knows that there
exists a critical curve $\sigma_0$  of (\ref{energy}), neither the
causal character of $\sigma_0$ nor (if the curve were timelike)
the time of arrival, would be known a priori.

In the case of the LFE, one can also consider a functional,
introduced in \cite{benci98}, which is related to action
functional and closer to (\ref{energy}):
\begin{equation} \label{j}
J_{x_{0},x_{1}}[\sigma]=\int _{0}^{1} \left(\frac{1}{2}
g(\sigma'(\lambda),\sigma'(\lambda))  +b\,
\omega(\sigma'(\lambda))\right)\dd \lambda ,
\end{equation}
on the space of all the (absolutely continuous) curves, non
necessarily causal, which connect $x_0$ and $x_1$ in the interval
$[0,1]$.  Concretely, Bartolo and Antonacci et al.
\cite{antonacci00,bartolo99} studied the connectedness of the
whole spacetime by means of critical points (non-necessarily
causal) of this functional, and further results were obtained in
posterior references (see, for example,
\cite{bartolo01,caponio02,caponio04,caponio04c,mirenghi02}
or the detailed acount in \cite{minguzzi04b}). Remarkably, in
\cite{caponio04d,caponio04b} the authors were able to prove, under
global hyperbolicity, the existence of at least one ({\em
uncontrolled}) value of $q/m$ such that a timelike connecting
solution of the associated LFE does exist. Indeed, a timelike
critical point $\sigma_0$ (if it exists) is a solution of the
Lorentz force equation {\em for some (uncontrolled) ratio $q/m$}.
This follows from the Euler-Lagrange equation for $J_{x_0,x_1}$
\begin{equation} \label{lorentz2}
 D_{\lambda} \left(\sigma'\right)=b\hat F(x)\left[\sigma' \right].
\end{equation}
In particular, $\dd s/\dd \lambda=C$ is a constant and, thus,
$\sigma_0$ would satisfy the LFE with charge-to-mass ratio
$q/m=b/C$. However, $C$ depends on the critical curve $\sigma_0$
($C=\int_{\sigma_0} \dd s$), which in turn depends on the
coefficient $b$ in an uncontrolled way.

Summing up, the  variational approach for functional
$J_{x_0,x_1}$, even though mathematically appropriate to study
non-timelike curves, presents the following two limitations for
our question (Q):
\begin{itemize}
\item[(a)] For chronologically related points, one cannot
control easily the causal character of the critical points.
\item[(b)] Even if the critical point is proved to be timelike,
one cannot know a priori its charge-to-mass ratio and indeed it
could in the end correspond to an unphysical value.
\end{itemize}

\subsection{Known results on (Q), and a counterexample}\label{ss33}

An approach conceived to study directly the physical question (Q)
started in \cite{caponio03}, \cite{minguzzi03b}, where the authors
considered the solutions to the LFE as projections of suitable
lightlike
geodesics for a Kaluza-Klein metric. 
It is well known the relation between the LFE and the projections
of timelike geodesics for a higher dimensional Kaluza-Klein
spacetime
\cite{kaluza21,kerner68,kerner01,kovacs84,leibowitz73,lichnerowicz55}.
However, it proved more useful for question (Q) to consider  the
solutions of the LFE as projections of {\em lightlike} geodesics
in a Kaluza-Klein spacetime with a scale factor (for the
additional dimension) proportional to the charge-to-mass ratio.
Indeed this allowed to prove the existence of solutions having an
a
priori fixed charge-to-mass ratio in most cases.

The results in \cite{minguzzi03b} improve those in
\cite{caponio03}; the best achieved result is then:
\begin{theorem}\label{mingu} 
Let $(M,g)$ be a globally hyperbolic spacetime, and $\eh=\dd \ptl$
be an 
electromagnetic field   on $M$. Let $x_{1}$ be an
event in the chronological future of $x_{0}$ and $q/m \in \R-\{ 0
\}$ any charge-to-mass ratio. Then there exists a future-directed
causal  curve $\sigma_0$ which connects $x_0$ and $x_1$ and
maximizes the functional $I_{x_0,x_1}$ on $\mathcal{N}_{x_0,x_1}$.

Moreover, $\sigma_0$ is everywhere timelike or lightlike. In the
former case, the reparametrization of $\sigma_0$ with respect to
proper time becomes a solution of the LFE (\ref{lorentz}); in the
latter case, $\sigma_0$ is a lightlike geodesic.
\end{theorem}

\noindent Even though the hypotheses of this theorem are optimal,
it is not completely satisfactory for our  question (Q) because
Theorem \ref{mingu} does not forbid the maximizing curve
$\sigma_0$ to be a  lightlike geodesic. We present below an
example of such a situation, even in a simply connected (and
contractible) spacetime. Summing up: {\em if a connecting
lightlike geodesic exists, Theorem \ref{mingu} does not answer our
question (Q)}.

\begin{example} \label{exribbon}{
Consider the 3-dimensional spacetime $M= \mathbb{R} \times S$,
$\dd s^2= \dd t^2 - \dd l^2$, $t \in \mathbb{R}$, in Example
\ref{ejecut}, Remark \ref{rch2}. Let $F=d\omega$ independent of
$t$, with $\omega\equiv 0$ in the cap of $S$ and, on the cylinder:
$\omega =B r \mu(z) \dd \theta$, where $\theta$ is the angle, $B
\in \mathbb{R}$ and $\mu(z)$ is a smooth monotone decreasing
function such that $\mu(-2\pi r)=1$, $\mu(-\pi r)=0$. Notice that
$F$ is different to zero only in the set $\mathbb{R} \times R$,
where the ribbon $R \subset S$ is defined by $R=\{q \in S : -2\pi
r < z < -\pi r\}$.

Let $p \in S$ be defined by the $\mathbb{R}^3$ coordinates
$(r,0,-3\pi r)$, and let $x_{0}=(0,p)$, $x_1=(2 \pi r, p)$. The
events $x_0$ and $x_1$ are connected by two lightlike geodesics
$\sigma_1$, $\sigma_2$. Their projections $\bar c$, $\tilde
c$ differ only on the orientation of their parametrization as
their image is the circle $x^{2}+y^2=r^2$, $z=-3\pi r$.
Let $\sigma_L$ be a generic timelike connecting curve such that
its projection $c_L$ on $S$ has length $L$. Since $\sigma_L$ is
timelike, $\dd l/\dd t < 1$, and since it is also connecting we
have $L< 2 \pi r$. This equation implies that any timelike
connecting curve has a projection $c$ completely contained in the
region $z \le - 2 \pi r$  ($c$ belongs to the identity of
the homotopy group of the cylindric part of $S$,
with base point $p$).  In particular timelike connecting curves
cannot enter the region  of non-vanishing electromagnetic field.

Thus, there are three causal homotopy classes. Two classes ${\cal C}_i$ for the
two isolated lightlike geodesics $\sigma_i$, $i=1,2$ and a
further causal class ${\cal C}$ containing all the timelike connecting
curves (this last class contains lighlike curves, which are not geodesics).
Roughly speaking the curves in ${\cal C}$ cannot be
deformed to the lightlike geodesics $\sigma_i$ since the projections would reach
the cap, and  hence the deformed curves would become non-causal.

We shall now show that, for any $\vert \frac{q}{m}B\vert > 1$, the
absolute maximum and minimum of $I_{x_0, x_1}[\sigma]$ are reached
in the geodesics $\sigma_i$. First, notice that the
electromagnetic term of the action can be rewritten
\begin{equation}
 \frac{q}{m}\int_{\sigma} \omega  =  \frac{q}{m} B r \oint_{c} \mu(z)
\dd \theta ,
\end{equation}
which vanishes on ${\cal C}$, and is equal to $\pm 2 \pi
\frac{q}{m} B r$ on the geodesics $\sigma_i$. Thus, on the  class
${\cal C}$ the action functional is equivalent to the length
functional, which is bounded by the length $2\pi r$ of the
maximizing geodesic (the particle at rest at $p$), as required.
}\end{example}

\section{Maximization of $I_{x_0,x_1}$ over homotopy classes} \label{s4}

First, let us introduce a trivial principal bundle $P=M \times
\mathbb{R}$, $\Pi : P \to M$ with structural group $(\R , +)$: $ b
\in \R$, $p=(x,y)$, $p'=p b=(x, y+b)$. Let $y$ be the fiber
coordinate and
 $\beta$ be a dimensional positive constant. Given a potential 1-form $\ptl$
on $M$
 and a connection $\tilde\ptl=\dd y + \beta
\ptl$ on $P$, consider the  Kaluza-Klein metric
\begin{equation}
\tilde g =g-a^{2}  \tilde\omega^{2}  \label{kk},
\end{equation}
and choose the scale factor $a$ as $a=\beta^{-1} \vert \frac{q}{m}
\vert$. The actual value of the dimensional constant $\beta$ will have no role in our
work\footnote{It should be said that in the physical Kaluza-Klein
theory one usually chooses $a= \beta^{-1}\sqrt{16 \pi G} $ where
$G$ is the Newton constant, to obtain the correct coupling between
gravity and electromagnetism. This choice is obviously
incompatible with our constraint (even approximately, because both
imply $\vert \frac{q}{m}\vert \frac{1}{\sqrt{16 \pi G}}=1$ and for
realistic particles this coefficient is huge). However, note that
the Kaluza-Klein spacetime is used by us only as a technical tool:
given $q/m$ we define $a$, and so $a$ changes with the particle
considered - the spacetime $P$ is not pretended to be a physical
spacetime.}.

Fix $x_0 \leq x_1$ and causal homotopy class ${\cal
C}_{x_0,x_1}$ of ${\cal N}_{x_0,x_1}$. We are looking for critical curves of (\ref{action})
on ${\cal C}_{x_0,x_1}$ (and, thus, on ${\cal N}_{x_0,x_1}$).

\subsection{Causal homotopy classes in a K-K bundle} \label{ss41}

Let   ${\cal C}_{p_0,x_1}$ denote a $C^1$ homotopy class of causal
curves on $P$,  starting at some $p_{0} \in \Pi^{-1}(x_{0})$ and
ending  in $\Pi^{-1}(x_{1})$ where, now, the homotopy does not
keep fixed the second endpoint.

\begin{proposition} \label{p4.2}
Fixed $x_0\leq x_1$ and $p_0\in \Pi^{-1}(x_0)$:
\begin{itemize}
 \item[(1)] If $ \gamma_1,
\gamma_2 \in {\cal C}_{p_0,x_1}$ then $\sigma_1= \Pi \circ
\gamma_1$ and $\sigma_2= \Pi \circ \gamma_2$
belong to the same class ${\cal C}_{x_0,x_1}$.
\item[(2)]  If $\gamma_1$ and $\gamma_2$ projects on curves
$\sigma_1= \Pi \circ  \gamma_1$ and $\sigma_2= \Pi \circ \gamma_2$
which are causally homotopic, then
 $\gamma_1$ and $\gamma_2$ belong to the same ${\cal C}_{p_0,x_1}$.
\end{itemize}
Thus, the projection $\pi: P \to M$ sends homotopy classes of type
${\cal C}_{p_0,x_1}$ to homotopy classes of type ${\cal
C}_{x_0,x_1}$, and induces a bijective map between classes type
${\cal C}_{p_0,x_1}$ and ${\cal C}_{x_0,x_1}$.
\end{proposition}

\begin{proof} Assertion (1)  is obvious. Recall also that it ensures the induction of
a map between homotopy classes. The surjectivity of this map is
ensured because, for any future-directed causal curve $\sigma$ in
$M$ which connects $x_0$ and $x_1$, there exists a future directed
causal curve $\tilde \sigma$ in $P$ (i.e., the horizontal lift)
starting  at $p_0$ and projecting on $ \sigma$.

For (2) (which ensures injectivity), consider first the case when
the $\gamma_i$'s project on the same curve $\sigma$, that is,
$$ \gamma_1(\lambda)=(\sigma(\lambda), y_1(\lambda)), \quad
\gamma_2(\lambda)=(\sigma(\lambda), y_2(\lambda)), \quad \forall
\lambda \in [0,1]. $$ The map
\begin{equation}
H(\epsilon,\lambda)=\left(\sigma(\lambda), \epsilon
y_1(\lambda)+(1-\epsilon) y_2(\lambda) \right).
\end{equation}
is a causal homotopy without the second point
fixed, as required.

If the projections $\sigma_i$ are different, consider their
horizontal lifts $\tilde \sigma_i$ starting at $p_0$. Then each
$\tilde \sigma_i$  is continuously $C^1$ causally homotopic to
$\gamma_i$  (from the previous case), and the lifting of the $C^1$
causal homotopy between $\sigma_1$ and $\sigma_2$ is clearly a
$C^1$ causal homotopy between $\tilde \sigma_1$ and $\tilde
\sigma_2$. \qed \end{proof}

\noindent \begin{remark} \label{rhomfib} {\rm An analogous result
holds for timelike classes type $C_{p_0,x_1}$. In general, all the
study of Subsection \ref{ss23} for the spaces containing piecewise
geodesics
 ${\cal M}_{x_0,x_1}$, $M_{x_0,x_1}$, and its arc-connected components
 ${\cal U}_{x_0,x_1}$, $U_{x_0,x_1}$
(constructed from causal and time path spaces from $x_0$ to
$x_1$), can be extended to analogous spaces containing piecewise
geodesics from $p_0$ to $\Pi^{-1}(x_1)$, namely, ${\cal
M}_{p_0,x_1}$, $M_{p_0,x_1}$, and its arc-connected components
${\cal U}_{p_0,x_1}$, $U_{p_0,x_1}$. Just take into account:

(1) If $t$ is a Cauchy temporal function on $M$ then so is $\tilde
t= t \circ \Pi$ on $P$ (see Appendix). In particular, $J^+(p_0)
\cap \Pi^{-1}(x_1)$  is compact because, chosen any Cauchy
hypersurface $S$ of $M$ through $x_1$, it can be written as
$\left(J^+(p_0) \cap \Pi^{-1}(S)\right) \cap \Pi^{-1}(x_1)$ (the
intersection between a compact and a closed subset).

(2) An analogous to Theorem \ref{tdelfin}, Remark \ref{rdelfin}
can be stated and, in particular, the limit of geodesics in a
class ${\cal C}_{p_0,x_1}$ (resp. $C_{p_0,x_1}$) also belong to
this class (resp. to the closure of this class). }
\end{remark}

\subsection{A Fermat-type equivalent problem} \label{ss42}

\noindent Given $\sigma: [0,1] \to M$ in ${\cal C}_{x_0,x_1}$,
consider the two lightlike lifts $$\tilde \sigma^\pm (\lambda) =
(\sigma(\lambda), y^\pm(\lambda)), \quad \lambda \in [0,1]$$ of
$\sigma$ in $P$ starting at a fixed $p_{0}=(x_{0}, y_{0}) \in
\Pi^{-1}(x_0)$.  Explicitly, the requirement to be lightlike
implies \be \label{yo0} (y^\pm) '(\lambda)=  \mp \frac{1}{a}
|\sigma'(\lambda)| -\beta \omega(\sigma'(\lambda )) \ee and, thus:
\begin{equation}
\tilde{\sigma}^{\pm}(\lambda)= \left(\sigma(\lambda),\, y_0 \mp
\frac{1}{a} \int_{0}^{\lambda}\sqrt{g(\sigma'(\lambda ),
\sigma'(\lambda))} \dd \lambda -\beta \int_{0 }^{\lambda}\omega
(\sigma'(\lambda)) \dd \lambda \right) .
\end{equation}
Then, the fiber coordinate $Y_{1}^{\pm}[\sigma]$ of the final point of this curve is, essentially, the
action functional:
\begin{equation} \label{ciao}
Y_{1}^{\pm}[\sigma]= y^{\pm}(1)=y_{0}\mp \frac{1}{a}\left(\int_{\sigma}
\dd s +(\pm \vert q/m\vert)  \int_{\sigma} \omega \right).
\end{equation}
Comparing this expression with the one of $I_{x_{0},x_{1}}$, it is
obvious that a maximization on ${\cal C}_{x_0,x_1}$ of
$I_{x_{0},x_{1}}$ relative to the ratio $+ \vert q/m \vert$ (resp.
$- \vert q/m \vert$), corresponds to  a minimization (resp.
maximization) of $Y_{1}^{+}[\sigma ]$ (resp. $Y_{1}^{-}[\sigma
]$). Summing up:

\bt \label{t} The curve $\sigma_0$ is a maximum of $I_{x_0,x_1}$
on ${\cal C}_{x_0,x_1}$ and $q/m>0$ (resp. $q/m<0$) if and only if
$\sigma_0$ is a minimum (resp. maximum) of the arrival coordinate
$$Y^+_1: {\cal C}_{x_0,x_1}\rightarrow \R , \quad \sigma \mapsto Y^+_1(\sigma)$$
$$\mbox{(resp.} \; Y^-_1: {\cal C}_{x_0,x_1}\rightarrow \R , \quad \sigma \mapsto Y^-_1(\sigma) \mbox{).}$$
\et

\subsection{The maximization result} \label{ss43}

The above variational principle reduces our problem to ensure the
existence of maxima or minima for $Y_1^\pm$. The following result
yields two candidates.

\begin{lemma} \label{l4.4}
The set $K$ containing the points in $\Pi^{-1}(x_1)$ reached by means of a causal curve in
${\cal C}_{p_0,x_1}$, is a compact interval $K= \{x_1\} \times [\underline{y}_1,  \bar y_1],
\underline{y}_1 \leq  \bar y_1$.

Moreover, $p_0$ will be connectable with any of the two extremes
of $K$ by means of a lightlike geodesic $\gamma^\pm$ (necessarily
without conjugate points before the endpoint) in ${\cal
C}_{p_0,x_1}$.
\end{lemma}
\begin{proof} The arc-connectedness of $K$ is straightforward from the
existence of a causal homotopy between any two curves in ${\cal
C}_{p_0,x_1}$. For the compactness of $K$, recall that as
$J^+(p_0) \cap \Pi^{-1}(x_1)$  is compact (Remark \ref{rhomfib},
item (1)),  any Cauchy sequence $\{(x_1,y_k)\}$ in $K$ will have a
limit $(x_1, y_\infty)$ in $\Pi^{-1}(x_1)$. By Avez-Seifert
theorem, there exists a maximizing causal geodesic $\gamma_k\in
{\cal U}_{p_0,x_1}\subset {\cal C}_{p_0,x_1}$ connecting $p_0$ and
each $y_k$. Then, the limit curve $\gamma_0$ of the sequence $\{
\gamma_k \}$ will be   a causal geodesic in ${\cal C}_{p_0,x_1}$
too, and it will cross the Cauchy hypersurface $\Pi^{-1}(S)$ at
some point, necessarily $(x_1, y_\infty)$. Thus, $\gamma_0$
belongs to the same causal homotopy class (Remark \ref{rhomfib},
item (2)).

For the last assertion, notice that the extreme $\underline{y}_1$
(resp. $\bar y_1$) is connectable with $p_0$ by means of a
length-maximizing causal geodesic $\gamma^+$ (resp. $\gamma^-$).
Even more, $\gamma^\pm$ must be lightlike because, otherwise, an
open neighborhood of the final point of $\gamma^\pm$ in
$\Pi^{-1}(x_1)$ would lie in $K$.
Finally,  a conjugate point cannot exist because $\gamma^\pm$ is
maximizing (see Theorem \ref{cch2}). \qed \end{proof}

\begin{lemma} \label{l4.5}
Assume $q/m > 0$ (resp. $<0$), and let $\gamma(\lambda) =
(\sigma(\lambda), y(\lambda)), \lambda\in [0,1]$ be the lightlike
geodesic in ${\cal C}_{p_0,x_1}$  which connects $p_0$ and $(x_1,
\underline{y}_1)$ (resp. $(x_1, \bar y_1)$). Then, $\sigma$ is a
maximum  of $I_{x_0,x_1}$ on ${\cal C}_{x_0,x_1}$.
\end{lemma}
\begin{proof} From Theorem \ref{t}, we only have to prove  $\gamma =
\tilde \sigma^+$ (resp $=\tilde \sigma^-$) because in this case
$\sigma$ is obviously a minimum (resp. maximum) of $Y^+$ (resp.
$Y^-$). As $\gamma$ is a geodesic and $\partial_y$ a Killing field
on $P$, we have the constant $\nu \equiv \tilde
g(\gamma',\partial_y) = -a^2\left( y' + \beta \omega
(\sigma')\right)$, that is: \be \label{yo1} y' =
-\frac{\nu}{a^2}-\beta\omega (\sigma'). \ee Even more, as the
extreme $(x_1, \underline{y}_1)$ is minimum (resp. $(x_1, \bar
y_1)$ maximum) in $K$, then $\nu \geq 0$ (resp. $\leq 0$) --
otherwise, the horizontal lift of $\sigma$ would end beyond the
extreme. As $\gamma$ is lightlike, $\tilde
g(\gamma',\gamma')\equiv 0$ thus \be \label{yo2} y' = \varepsilon
\frac{|\sigma'|}{a}-\beta\omega (\sigma ') \ee  where
$\varepsilon$ equals $1$, or $-1$. To specify $\varepsilon$,
notice from (\ref{yo1}), (\ref{yo2}) and the sign of $\nu$: \be
\label{yo3} |\sigma'| = + \frac{\nu}{a} \quad \quad \mbox{(resp. }
|\sigma'| = - \frac{\nu}{a} \mbox{)} \ee Now, use (\ref{yo3}) to
check that the expression for $(y^+)'$ (resp. $(y^-)'$) in
(\ref{yo0}) coincides with the expression of $y'$ in (\ref{yo1}),
and the result follows. \qed \end{proof}

\smallskip

\noindent Notice also that, in the proof, $\nu = 0$ if and only if
$\sigma $ is a lightlike geodesic and $\gamma$ is its horizontal
lift \cite{minguzzi03b}; otherwise $\sigma$ is timelike.

\bere { Analogous results hold if $K$ in Lemma \ref{l4.4} is taken
as the adherence of the points in $\Pi^{-1}(x_1)$ reachable by
curves in a timelike homotopy class $C_{p_0,x_1}$. The lighlike
geodesics $\gamma^\pm$  will lie in  $\overline{C}_{p_0,x_1}$, and
will be causally homotopic to the curves in $C_{p_0,x_1}$. The
analog of  Lemma \ref{l4.5} states that if $q/m>0$ (resp.
$q/m<0$), $\Pi(\gamma^{+})$ (resp. $\Pi(\gamma^{-})$) maximizes
$I_{x_0,x_1}$ on $\overline{C}_{x_0,x_1}$.  This projection is
either a null geodesic belonging to the boundary
$\dot{C}_{x_0,x_1}$ or a timelike curve belonging  to
${C}_{x_0,x_1}$. }\ere
Summing up, 
the following generalization of Theorem \ref{mingu} to
causal homotopy classes is obtained:

\bt \label{yomain}
Let $(M,g)$ be a globally hyperbolic spacetime, and $\eh=\dd \ptl$ be an
electromagnetic field   on $M$. Let $x_{1}$ be an
event in the causal future of $x_{0}$ and fix  any  causal homotopy class
${\cal C}_{x_0,x_1}$.

For each $q/m \in \R-\{ 0\}$ there exists a future-directed causal  curve $\sigma_0$
which connects $x_0$ and $x_1$
and maximizes the functional $I_{x_0,x_1}$ on
$\mathcal{C}_{x_0,x_1}$.

Moreover, $\sigma_0$ is everywhere timelike or lightlike. In the
former case, the reparametrization of $\sigma_0$ with respect to
proper time becomes a solution of the LFE (\ref{lorentz}) for the
charge-to-mass ratio $q/m$; in the latter case, $\sigma_0$ is a
lightlike geodesic.

Even more,  for any timelike homotopy class $C_{x_0,x_1} \subset
{\cal N}_{x_0,x_1}$ there exists a maximizer in
$\overline{C}_{x_0,x_1}$ which is either a timelike curve in
$C_{x_0,x_1}$  or a lightlike geodesic in the boundary
$\dot{C}_{x_0,x_1}$.


\et

\section{Existence of timelike local maximizers} \label{s5}

\subsection{The existence
result for causal homotopy classes.} \label{s5.1} In the previous section the existence of maximizers, either
timelike curves or lightlike geodesics, in each ${\cal C}_{x_0,x_1}$
($\overline{C}_{x_0,x_1}$), has been ensured. Here, we will prove
 that the maximizer on ${\cal C}_{x_0,x_1}$ cannot be a lightlike geodesic if there exists
a timelike curve in ${\cal C}_{x_0,x_1}$. The proof is carried out
in two steps. In the first one (Lemma \ref{yolema}), a maximizing
lightlike geodesic $\sigma_0$ is shown to be free of conjugate
points (except at most the two extremes). The second step is  to
check that, if such a $\sigma_0$ exists, all the other curves in
${\cal C}_{x_0,x_1}$ must be lightlike.

\begin{lemma}\label{yolema}
Let $\sigma: [0,1] \rightarrow M$ be a lightlike geodesic such
that $\sigma (r), 0<r<1$ is the first conjugate point to $\sigma
(0)$. Then there exists a  smooth ($C^{r_0}$) variation of
$\sigma$ through causal curves such that the functional
$I_{x_0,x_1}$ is strictly bigger on the variated longitudinal
curves.

Therefore, the maximum of $I_{x_0,x_1}$ on a causal homotopy class
${\cal C}_{x_0,x_1}$  cannot be attained at a lightlike geodesic
with a conjugate point to $\sigma(0)=x_0$ before $\sigma(1)=x_1$.
\end{lemma}

\begin{proof} The integrand in the right hand side of
(\ref{action}) will be written, for a general smooth variation
through causal curves $\sigma_v(\lambda)$, $v\in [0,\epsilon ],
\epsilon >0$, as \be \label{estrella}
 \tilde I\equiv  \tilde I_v(\lambda)= \sqrt{\langle \sigma_v'(\lambda), \sigma_v'(\lambda) \rangle}
 + \frac{q}{m} \omega(\sigma_v'(\lambda)).
\ee
 Let $V(\lambda)$ be the variational field. Recall that for this variation, at $v=0$:
$$\langle \sigma_v'(\lambda), \sigma_v'(\lambda) \rangle \equiv 0 \quad \partial_v
\langle \sigma_v'(\lambda), \sigma_v'(\lambda) \rangle =
2 \langle V(\lambda), \sigma_v'(\lambda) \rangle \equiv 0 .$$ But
the second derivative of $\langle \sigma_v'(\lambda),
\sigma_v'(\lambda) \rangle$ at $v=0$ can be chosen nonnegative on
$[0,1]$ and strictly positive in some interval $(0,r+\delta)$ as
in \cite[Proposition 10.48]{oneill83}. Moreover, this second
derivative is equal for the associated  variation
$\sigma_{-v}(\lambda)$ with variational field $-V$. Then, by using
a Taylor expansion of (\ref{estrella}): \be \label{1}
 \frac{d\tilde I}{dv}(\lambda)\mid_{v=0}=
\sqrt{\frac{\partial^2}{2\partial v^2} \langle \sigma_v'(\lambda),
\sigma_v'(\lambda) \rangle \mid_{v=0}}+ \frac{q}{m}
\partial_v(\omega (\sigma_v'(\lambda)))\mid_{v=0} , \quad \forall
\lambda \in (0,r+\delta), \ee where the integral $\int_0^1$ of the
first term is strictly positive and equal for the two variations
$\sigma_v(\lambda)$ and $\sigma_{-v}(\lambda)$. As the integral of
the last term changes with the sign of $V$,  the integral of
(\ref{1}) will be strictly positive for at least one of the two
variations, as required. \qed \end{proof}

\begin{remark} \label{r5.2} {
Even though in Lemma \ref{l4.4} the obtained lightlike geodesic in
the total space $P$ cannot have a conjugate point, we have proved
here directly the inexistence of conjugate points for its
projection on $M$. In fact, the proof shows that this is a general
property for actions type $I_{x_0,x_1}$, which contain a free
particle term plus lower order terms in $|\sigma'|$. }\end{remark}

With Lemma \ref{yolema} at hand, the last step follows just
applying the studied properties of causal homotopy classes.

\bt \label{yomain2} Under the hypotheses of Theorem \ref{yomain},
 the maximizer $\sigma_0$ of $I_{x_0,x_1}$ on ${\cal C}_{x_0,x_1}$  is timelike if ${\cal C}_{x_0,x_1}$ contains a timelike
curve. \et

\begin{proof} Assume by contradiction that $\sigma_0$ is not timelike and,
thus, it is  a lightlike geodesic with no conjugate points before
$x_1$. By Theorem \ref{cch2}, $\sigma_0$ must maximize the time
separation in ${\cal C}_{x_0,x_1}$, in contradiction with the
existence of a timelike curve in ${\cal C}_{x_0,x_1}$. \qed
\end{proof}

Theorems \ref{yomain} and \ref{yomain2} prove directly our main result, Theorem \ref{tyomain}.

\subsection{A remarkable example} \label{s5.2} Lemma \ref{yolema} does not forbid the existence of a lightlike
geodesic $\sigma$ which maximizes the functional on the closure of
a timelike class $\overline{C}_{x_0,x_1}$. However, in that case
the maximizer on ${\cal C}_{x_0,x_1} \supset
\overline{C}_{x_0,x_1}$ does not coincide with $\sigma$, as the following example shows.

\begin{example} Let $\Sigma$ be a surface embedded in $\mathbb{R}^3$ obtained
by gluing the spherical cap $x^{2}+y^{2}+z^2=r^2$, $z>
-\frac{\sqrt{3}}{2} r + \epsilon_z$, with a cylinder
$x^{2}+y^2=r^2/4$, $z< -\frac{\sqrt{3}}{2} r- \epsilon_z$,  by
making a smooth transition in the points with coordinate $z \in
[-\frac{\sqrt{3}}{2} r- \epsilon_z , -\frac{\sqrt{3}}{2} r+
\epsilon_z]$, for some  positivive $\epsilon_z <
\frac{\sqrt{3}}{2} r$. Notice that this transition can be made
smooth and depending only on the azimuthal angle $\theta$ in a
small interval  $(\frac{5}{6}\pi - \epsilon_\theta, \frac{5}{6}\pi
+ \epsilon_\theta), \epsilon_\theta<\pi/6$. Only the details of
this surface included in the spherical cap with $\theta \le \pi/2
+\epsilon$, for some small positive $\epsilon<\pi/6$,  will be
relevant.

Let $d l^2$ be the induced Riemannian metric on $\Sigma$, and fix
$q=(r,0,0) \in \Sigma$. Consider the natural product (globally
hyperbolic) spacetime $M=\mathbb{R} \times \Sigma$, $g=dt^2-dl^2$,
with natural projection $\pi: M \to \Sigma$, and the fixed events
$x_0=(0,q)$, $x_1=(2\pi r,q)$. The timelike curve $\lambda \mapsto
(2 \pi r \lambda, q)$ fix a timelike homotopy class $C_1 (:=
C^{(1)}_{x_0,x_1})$. The connecting lightlike geodesic
$$\sigma_0(\lambda) = (2 \pi r \lambda, c_0(\lambda)), \quad c_0(\lambda) =(r \cos 2 \pi \lambda,r\sin
2 \pi \lambda,0), \quad \quad \lambda\in [0,1],$$ lies in the
boundary $\dot C_1$. In fact, $\sigma_0$ can be reached by
approximating the part $c_0$ with a constant-speed parametrization
$c_\alpha$ of $\Sigma \cap \Pi_\alpha$, where $\Pi_\alpha \subset
\mathbb{R}^3$ is the plane through $q$, orthogonal to the plane
$y=0$, which makes an oriented positive angle $\alpha < \pi/2$
with the plane $z=0$ ($c_\alpha$ is contained in the region $z>0$
except in the tangent point $q$). However, by letting $\alpha<0$
we can find a second timelike homotopy class $C_2$ such that
$\sigma_0 \in \dot C_2$; of course, $C_1$ and $C_2$ are contained
in the same causal homotopy class ${\cal C}$. Notice that $c_0$
passes through the antipodal point $-q=(-r,0,0)$, which is also a
conjugate point of $q$; thus, $\sigma_0$ also contains a conjugate
point.


Fix $q/m>0$ (resp. $q/m<0$), and let $F={\cal B }\pi^{*}\Omega =
d\omega$ be on $M$, where $\Omega$ is the volume 2-form of
$\Sigma$ (with the  orientation induced by the outer normal in the
spherical cap), and where ${\cal B}: \Sigma \to \mathbb{R}$ is a
non-negative (resp. non-positive) function, with  ${\cal B}\equiv
B>0$ (resp. $<0$) constant for $\theta \le \pi/2$, and
monotonically decreasing (resp. increasing) to $0$ for $\theta \in
(\pi/2,\pi/2+\epsilon]$. The charged-particle action $I_{x_0,x_1}$
is given by two contributions. The electromagnetic term reads
\begin{equation}
\frac{q}{m}\int_{\sigma} \omega= \frac{q}{m} \int_{R} {\cal B}
\Omega
\end{equation}
where, without loss of generality, $\sigma(\lambda)=(2 \pi
r\lambda, c(\lambda))$ and $\p R=c$.   For a given length $L \le 2
\pi r$ of $c$ this integral is maximized in $C_1$
by the circle $c_\alpha$ with length $L$, namely $c^{L}$.  
Indeed, the maximizer must be a
circle in order to maximize the area,  and  it is
tangent to $c_0$ since, otherwise, its enclosed surface $R$ would
include regions where ${\cal B} <B$ (resp. ${\cal B}
> B$). Thus
\begin{equation}
\left|  \frac{q}{m}\int_{\sigma}  \omega \right| \le 
\frac{q}{m}B 
A[c^L]
\end{equation}
where $A[c^L]$ is the area contained in $c^L$. And the
equality holds iff $c =c^L$ (up to a reparametrization with the same winding number).

The contribution of the length  of $\sigma$ in $I_{x_0,x_1}$ is:
\begin{equation}
\int_{\sigma} \dd s=\int_{0}^{2\pi r}\sqrt{1-\left(\frac{\dd
l}{\dd t}\right)^2} \dd t \le 2 \pi r
\sqrt{1-\left(\frac{l[c]}{2\pi r}\right)^2}
\end{equation}
where $l[c]$ is the length of $c=\pi\circ \sigma$, and the
equality holds when the speed of $c$ is constant. We have then
\begin{equation}
I_{x_0,x_1}[\sigma] \le 2 \pi r\sqrt{1-\left(\frac{l[c]}{2\pi
r}\right)^2}+
\frac{q}{m}B
A[c^{l[c]}]
\end{equation}
where the equality holds iff $\pi \circ \sigma=c^{l[c]}$. But in
terms of the angle $0 \le \alpha \le \pi/2$ with $c_\alpha=c^l$,
we have $l=2 \pi r \cos \alpha$ and $A[c^{l}]=2 \pi r^2 (1-\sin
\alpha)$. Hence if
$
\frac{q}{m}B
r>1$
\begin{equation}
I_{x_0,x_1}[\sigma] \le 2 \pi r^2 \frac{q}{m}B  + 2 \pi r(1-
\frac{q}{m}Br) \sin \alpha \le 2 \pi r^2 \frac{q}{m}B
=I_{x_0,x_1}[\sigma_0],
\end{equation}
and the equality holds iff $\alpha=0$ and the projection of
$\sigma$ is   $c_0 (=c^{2 \pi r})$,
i.e. iff $\sigma=\sigma_0$.

Nevertheless, even though $\sigma_0$ maximizes in
$\overline{C}_1$, it {\em does not maximize in $\overline{C}_2$}
(nor in the causal homotopy class ${\cal C}$), in agreement with
our results. \end{example}

\bere \label{rs5.2}  Take the surface $\Sigma =S$ obtained by gluing the semisphere $x^2+y^2+z^2=r^2, z\leq 0$ with the semicylinder $x^2 +y^2= r^2, z\leq 0$. By using an appropiate differentiable structure, $S$ can be regarded as a $C^2$ manifold endowed with a $C^1$ metric. Thus, Christoffel symbols and geodesics make sense, but not conjugate points or Jacobi fields.   Repeating the above procedure, the timelike homotopy class $C_1$ is defined as above, but $C_2$ will not exist and, then, ${\cal C}= \overline C_1$.
Analogous $F, \omega$ makes sense on $z\geq 0$ (and can be extended to $S$), thus, the action $I_{x_0,x_1}$ can be defined on ${\cal C}$. Then, {\em the lightlike geodesic $\sigma_0$ maximizes in ${\cal C}$, and no timelike maximizer in ${\cal C}$ exists, even though ${\cal C}$ contains timelike curves}.

\ere

\section{The non-exact electromagnetic field case} \label{nex}

Throughout the paper, $F$ has been not only a {\em closed}
skew-symmetric 2-covariant vector field (i.e., an `electromagnetic
field'), but also {\em exact}. This condition is scarcely
restrictive: from the mathematical viewpoint, it is fulfilled in
any contractible spacetime and, from the physical one, no known
experimental evidence of non--exact electromagnetic fields on
spacetime exists (magnetic monopoles have not been found).
Nevertheless, it is interesting to study this case, in order to
understand better our approach and the limits of the expected
results. Recall:
\begin{itemize}
\item[(1)] If $F$ is not exact the problem of maximizing the action
becomes ill posed, since the electromagnetic potential is not
globally defined. Thus, the variational problem of finding
maximizing curves for the action does not make sense, but the
existence problem  for connecting solutions of the LFE is still
perfectly meaningful.
\item[(2)] There are non-exact $F$ which can
be studied by means of Kaluza-Klein metrics. For such a $F$, the
total space $P$, whenever it exists, is necessarily a  non-trivial
principle bundle with fiber $U(1)\equiv S^1$. In fact, the
necessary and sufficient condition for $\beta F$ to be the
curvature 2-form of a real-valued connection $\tilde\ptl=\dd y +
\beta \ptl$ over a $U(1)$ bundle $P$ on $M$ of fiber angle $y$
(and, introducing a metric (\ref{kk}), of fiber circumference
$2\pi a$), is that the cohomology class $[\frac{\beta}{2\pi } F]
\in H^{2}(M,\mathbb{R})$ is integer (more precisely,  the
$\breve{C}$ech cohomology class canonically associated with
$[\frac{\beta}{2\pi} F]$ belongs to the image of the morphism
$\epsilon^{2}: \breve{H}^{2}(M,\mathbb{Z}) \to
\breve{H}^{2}(M,\mathbb{R})$ induced by the inclusion $\epsilon:
\mathbb{Z} \to \mathbb{R}$.)\footnote{Remarkably, in this case the
complex exponential associated to the charged particle action
$e^{\frac{i}{a}I}$, can be well defined. The problems for $I$ can
therefore be circumvented in the quantum  case \cite{alvarez85}.}
\end{itemize}
\smallskip
\smallskip

\noindent The example below shows that Theorem \ref{tyomain} does
not admit a generalization to the non-exact case, and becomes
definitive for several reasons: the spacetime  and the field $F$
are very simple,  $F$ can be described with a principal bundle,
and the found events  $x_0$, $x_1$ cannot be connected by a
solution of the LFE, whatever the value of $q/m$ is chosen.

\begin{example} {
Let the 3-dimensional spacetime $M$ be the product $M=\mathbb{R}
\times S^2$, $t \in \mathbb{R}$,  with natural projection $\pi: M \to S^2$,
and metric $\dd s^{2}=\dd t^2-
\dd l^2$ where $\dd l^2$ is the usual Riemannian metric of a sphere
$S^2$ of radius $r$. 
Let $\Omega$ be one of the two associated volume 2-forms on $S^2$,
with associated endomorphism field $\hat \Omega \equiv (\Omega_{\
i} ^k)$, and consider on $M$ the electromagnetic
field\footnote{Notice that the electric part of $F$ with respect
to $\partial_t$ is 0, and the magnetic part corresponds, at a
prerelativistic level, with a uniform magnetic field on $S$, see
\cite[Section 3]{barros04} and references therein.} $F=B
\pi^{*}\Omega$.
 The LFE for a curve $\sigma (t)=(t,c(t))$ becomes
\begin{equation}
 D_{t} \frac{c'}{\sqrt{1-|c'|^2}}=\frac{q}{m}B \,\hat \Omega [c']
\end{equation}
where 
$D$ is the Levi-Civita connection of
$S^2$. Multiplying by $c'/\sqrt{1-|c'|^2}$ one finds $|c'|^{2}=const.$
Thus, from the above equation:
\begin{equation} \label{fre}
D_{t} c' =\frac{q}{m}B\,\sqrt{1-|c'|^2} \,\hat\Omega [c'],
\end{equation}
from which it follows that $c$
is a curve having constant velocity and
constant curvature. Indeed, the equation above implies that the
velocity $c'(t)$ 
rotates with angular velocity $\omega$,
$\vert \omega \vert= \vert \frac{q}{m}B \vert  \sqrt{1-|c'|^2}$ with
respect to a parallel transported frame along $c$. But the only
constant curvature curves on $S^2$ are the circles, and $c$ cannot be a maximal circle (in this case $c$ would be a geodesic and, hence, the left-hand side of (\ref{fre}) would vanish,
whereas the right-hand side would not). As a consequence, there is
no solution $c$ of (\ref{fre}) which connects two opposite points $p$, $q$ on $S^2$,
although for $t>\pi r$,  $x_1=(q,t)$ lies in the chronological
future of $x_{0}=(p,0)$.

}\end{example}


\section{Conclusions}\label{s6}

A full Avez-Seifert type result on the spacetime connectedness
through  solutions to the LFE has been proved. The hypothesis  of
this result (Theorem \ref{tyomain}) are optimal because, on one
hand, no more general hypothesis than $x_0 \ll x_1$ makes sense
and, on the other, the generalization to non-exact electromagnetic
fields is not possible.

 The proof is based in a
purely geometric technique on an auxiliary Kaluza-Klein spacetime,
and the consistency of the technique is proved by using the
variational interpretation of the solutions to the LFE for exact
electromagnetic fields. However, the proof itself is not
variational, and the results previously obtained by means of
variational methods  do not have the accuracy and natural physical
interpretation of those studied here.

By the way, new properties of timelike and causal homotopy classes, interesting
on their own right, have been obtained.
The careful study of the lightlike geodesics  in such classes have
become essential for our proof, and results as Theorems \ref{tch},
\ref{cch2} do not have analogous in the Riemannian case. Thus, the
applications of timelike and causal classes in the present and previous papers, as
\cite{galloway84,sanchez05,smith67}, would justify a further
study, as a separate field.

\section*{Appendix: causality of Kaluza-Klein metrics} \label{sa1}

Let $(M,g)$ be a spacetime and $G$ a Lie group endowed with a
positive definite Ad-invariant metric $h$. Let $\Pi: P\rightarrow
M$ be a principal fiber bundle with structural group $G$ endowed
with a Kaluza-Klein metric $$\tilde g = \Pi^* g - \tilde\omega^* h
,$$ where $\tilde\omega$ is some fixed 1-form connection on $P$.

Recall that, when $(M,g)$ is globally hyperbolic  then it admits a
Cauchy temporal function $t$; that is, $t$ is smooth with
future-directed timelike gradient\footnote{This means that $t$ is
temporal; in particular, $t$ is a time function, that is, a
continuous function which grows on any future-directed causal
curve.} and each level $S_{t_0}=t^{-1}(t_0)$ is a Cauchy
hypersurface \cite{bernal04}. Moreover, $M$ splits smoothly as a
product $\R \times S$ where $S$ is a Cauchy hypersurface and the
metric has no crossed terms between $\R$ and $S$.

\bt If $(M,g)$ is globally hyperbolic then $(P,\tilde g )$ is
globally hyperbolic too and, for any Cauchy temporal function $t:
M\rightarrow \R$ the composition $\tilde t= t\circ \Pi:
P\rightarrow \R$ is Cauchy temporal on $P$.

Therefore, $P$ splits smoothly as a product $\Pi^{-1}(S)\times \R$
where the  metric $\tilde g$ has no crossed terms and
$\Pi^{-1}(S)$ is a principal fiber bundle on $S$ with structural
group $G$. \et

\begin{proof} All the conclusions follow easily by proving that $\tilde t$
is a Cauchy temporal function. Clearly
$$ \tilde g (\nabla \tilde t, \tilde V) = d\tilde t(\tilde V) = dt (d\Pi(\tilde V)) = g(\nabla t, d\Pi(\tilde V)).$$
Thus, $\nabla \tilde t$ is the horizontal lifting of $\nabla t$ and, in particular, a temporal function.

To check that each hypersurface $\tilde S_{t}= \tilde t^{-1}(t)=
\Pi^{-1}(S_{t})$ is Cauchy, let   $\gamma$ be an inextendible
timelike curve, which can be assumed to be reparametrized with
$\tilde t$ without loss of generality. Assume by contradiction
that $\gamma$ crosses the hypersurfaces $\tilde S_t$ for $t<t_0$
but not $\tilde S_{t_0}$ (an analogous reasoning holds for $t_0<
t$). The projection $\sigma = \Pi \circ
 \gamma$ is also timelike, but it is extendible through $t_0$
because $S_{t_0}$ is Cauchy.  Then, consider a local
trivialization $U\times G$ of the bundle $P$ with $\sigma(t_0)\in
U$. Then we can write in this trivialization $\gamma (t) =
(\sigma(t), \eta(t))$ and
\begin{equation}
0<\tilde g(\gamma'(t),\gamma'(t)) = g(\sigma'(t),\sigma'(t)) -
h(\tilde\omega(\eta'(t)), \tilde\omega(\eta'(t))).
\end{equation}
That is, the $h$-length of $\eta(t)$ is bounded in $[t_0-\epsilon,
t_0)$ by the $g-$length of the extension of $\sigma$ to
$[t_0-\epsilon, t_0]$. Thus, as $h$ is complete, $\eta(t)$ is
continuously extendible to $t_0$, and so is $\gamma$, a
contradiction. \qed \end{proof}

\smallskip

\begin{remark} { In general, causality properties on $M$ (as  being chronological,
causal, strongly causal or stably causal) are transferred to $P$.
}\ere

\section*{Acknowledgements}

E.M. is supported by INFN, grant $\textrm{n}^{\circ}$ 9503/02, and
M.S. is partially supported by MEC-FEDER grant
$\textrm{n}^{\circ}$ MTM 2004-04934-C04-01.


\end{document}